\documentstyle[epsfig,a4p,11pt,cite]{article}

\newcommand{\PNxxx}    
    
\def\eplemi{\mbox{$\mathrm{e^+ e^-}$}}

\def\Zz{\mbox{$\mathrm{Z}^0$}}

\def\Pip{\mbox{$\mathrm{ \pi^+}$}}

\def\Km{\mbox{$\mathrm{ K^-}$}}

\def\Dz{\mbox{$\mathrm{ D^0}$}}

\def\Dstar{\mbox{$\mathrm{ D^{\star\pm}}$}}
\def\Dss{\mbox{$\mathrm{ D^{\star \star}}$}}

\def\Ups4s{\mbox{$\Upsilon(4S)$}}
\def\etal{\mbox{{\it et al.}}}
\def\ISGWSS{\mbox{ISGW$^{\star\star}$}}

\def\ra{\rightarrow}
\def\GeVcc{\mbox{GeV/$c^2$}}
\def\GeVc{\mbox{GeV/$c$}}
\def\GeV{\mbox{GeV}}
\def\MeVc{\mbox{MeV/$c$}}
\def\MeVcc{\mbox{MeV/$c^2$}}

\def\bD{\mbox{$\mathrm{b\ra D}$}}

\def\bcl{\mbox{$\mathrm{b \ra c \ra \ell}$}}

\newcommand{\inmath}[1] {\ifmmode#1\else$#1$\fi}
\newcommand{\definmath}[2] {\def#1{\ifmmode#2\else$#2$\fi}}
\newcommand{\alphas} {\alpha_{\mathrm{s}}}

\definmath{\dEdx} {{\mathrm d}E/{\mathrm d}x}
\definmath{\PWpm} {\mathrm{W}^{\pm}}      
\definmath{\Pgtp} {\tau^{+}}        
\definmath{\Pgtm} {\tau^{-}}        
\definmath{\Pgtpm}   {\tau^{\pm}}         
\definmath{\Pgn}  {\nu}          
\definmath{\Pagn} {\overline{\nu}}     
\definmath{\Pq}      {\mathrm{q}}
\definmath{\Paq}  {\overline{\mathrm{q}}}
\definmath{\PQ}      {\mathrm{Q}}
\definmath{\PaQ}  {\overline{\mathrm{Q}}}
\definmath{\Pu}      {\mathrm{u}}
\definmath{\Pau}  {\overline{\mathrm{u}}}
\definmath{\Pd}      {\mathrm{d}}
\definmath{\Pad}  {\overline{\mathrm{d}}}
\definmath{\Ps}      {\mathrm{s}}
\definmath{\Pas}  {\overline{\mathrm{s}}}
\definmath{\Pc}      {\mathrm{c}}
\definmath{\Pac}  {\overline{\mathrm{c}}}
\definmath{\Pb}      {\mathrm{b}}
\definmath{\Pab}  {\overline{\mathrm{b}}}
\definmath{\Pt}      {\mathrm{t}}
\definmath{\Pat}  {\overline{\mathrm{t}}}
\definmath{\Pap}  {\overline{\mathrm{p}}}
\definmath{\Pan}  {\overline{\mathrm{n}}}
\definmath{\PaD}  {\overline{\mathrm{D}}}
\definmath{\PaDz} {\overline{\mathrm{D}}^{0}}
\definmath{\PaB}  {\overline{\mathrm{B}}}
\definmath{\PaBz} {\overline{\mathrm{B}}^{0}}
\definmath{\PsDpm}   {\mathrm{D}^{\pm}_{\mathrm{s}}}  
\definmath{\PcgLpm}  {\Lambda^{\pm}_{\mathrm{c}}}  
\definmath{\PDst} {\mathrm{D}^{*}}     
\definmath{\PKs} {\mathrm{K}^{0}_{\mathrm s}}     
\definmath{\PgLz} {\Lambda^{0}}        
\newcommand{\qqbar}  {\Pq\Paq}

\newcommand{\qqg}  {\Pq\Paq g}

\newcommand{\ccbar}  {\Pc\Pac}
\newcommand{\bbbar}  {\Pb\Pab}

\newcommand{\Ztobb}     {\Zz\to\bbbar}
\newcommand{\Ztocc}     {\Zz\to\ccbar}

\newcommand{\Gammaof}[1]   {\Gamma_{\!\smash{#1}\mathstrut}}

\newcommand{\Gcc}    {\Gammaof{\ccbar}}
\newcommand{\Gbb}    {\Gamma_{\mathrm b \overline{\mathrm b}}}
\newcommand{\Ghad}      {\Gamma_{\mathrm{had}}}
\newcommand{\GbbGhad}      {\Gbb/\Ghad}
\newcommand{\GccGhad}      {\Gcc/\Ghad}

\newcommand{\costhe} {\cos\theta}

\definmath{\GeV}  {\mathrm{GeV}}
\definmath{\GeVc} {\mathrm{GeV}\!/c}
\definmath{\GeVcc}   {\mathrm{GeV}\!/c^2}
\definmath{\MeV}  {\mathrm{MeV}}
\definmath{\MeVc} {\mathrm{MeV}\!/c}
\definmath{\MeVcc}   {\mathrm{MeV}\!/c^2}
\definmath{\MVm}  {\mathrm{MV}\!/\mathrm{m}}
\definmath{\keV}  {\mathrm{keV}}
\definmath{\keVcm}   {\mathrm{keV}\!/\mathrm{cm}}
\definmath{\kV}      {\mathrm{kV}}
\definmath{\km}      {\mathrm{km}}
\definmath{\meter}   {\mathrm{m}}
\definmath{\cm}      {\mathrm{cm}}
\definmath{\mm}      {\mathrm{mm}}
\definmath{\micron}  {\mu\mathrm{m}}
\definmath{\nm}      {\mathrm{nm}}
\definmath{\kg}      {\mathrm{kg}}
\definmath{\gram} {\mathrm{g}}
\definmath{\second}  {\mathrm{s}}
\definmath{\microsec}   {\mu\mathrm{s}}
\definmath{\degree}  {^\circ}
\definmath{\degC} {^\circ\mathrm{C}}
\definmath{\ohm}  {\Omega}
\definmath{\Mohm} {\mathrm{M}\Omega}
\definmath{\rad}  {\mathrm{rad}}
\definmath{\mrad} {\mathrm{mrad}}
\definmath{\nb}      {\mathrm{nb}}
\newcommand{\eqref}[1]  {(\ref{#1})}
\newcommand{\PhysLett}  {Phys.~Lett.}

\newcommand{\NIM} {Nucl.~Instr.\ Meth.}
\newcommand{\ZPhys}  {Z.~Phys.}

\newcommand{\OPALColl}  {OPAL Collab.}
\newcommand{\JADEColl}  {JADE Collab.}

\def\cent{\centerline}

\def\th{\theta}
\def\to{$\rightarrow$}

\def\gcc{${\mathrm g}_{{\mathrm c}\overline
{\mathrm c}}$}
\def\gccd{{\mathrm g}_{{\mathrm c}\overline
{\mathrm c}}}
\def\gbb{${\mathrm g}_{{\mathrm b}\overline
{\mathrm b}}$}
\def\gbbd{{\mathrm g}_{{\mathrm b}\overline
{\mathrm b}}}

\def\plm{$\pm$}

\begin{document}
\begin{titlepage}
\begin{center}
{\Large EUROPEAN LABORATORY FOR PARTICLE PHYSICS}
\end{center}

\begin{flushright}
       CERN-EP/99-089   \\ July 1, 1999
\end{flushright}

\vspace{4cm}
 
\begin{center}
{\Huge\bf Measurement of the Production Rate of Charm Quark
  Pairs\\  from Gluons in Hadronic Z$^{\mathbf{0}}$ Decays}
\end{center}
\vspace{1cm}
\begin{center}
{\Huge The OPAL Collaboration}
\end{center} 
 
\vspace{1.5cm}
\cent{\large\bf Abstract}
\noindent
The rate of secondary 
charm-quark-pair production has been measured
in 4.4~million hadronic Z$^0$ decays 
collected by OPAL.
By selecting events with three jets and tagging 
charmed hadrons in the gluon jet candidate using leptons and \Dstar\
mesons, the average 
number of secondary charm-quark pairs per hadronic 
event is found to be (3.20\plm0.21\plm0.38)$\times10^{-2}$.

\vspace{1.5cm}

\cent {\large To be submitted to European Physical Journal C}
\vspace{.5cm}
\vspace{3cm}

\end{titlepage} 
 
\begin{center}{\Large        The OPAL Collaboration
}\end{center}\bigskip
\begin{center}{
G.\thinspace Abbiendi$^{  2}$,
K.\thinspace Ackerstaff$^{  8}$,
G.\thinspace Alexander$^{ 23}$,
J.\thinspace Allison$^{ 16}$,
K.J.\thinspace Anderson$^{  9}$,
S.\thinspace Anderson$^{ 12}$,
S.\thinspace Arcelli$^{ 17}$,
S.\thinspace Asai$^{ 24}$,
S.F.\thinspace Ashby$^{  1}$,
D.\thinspace Axen$^{ 29}$
% ...
}\end{center}\bigskip
\bigskip
$^{  1}$School of Physics and Astronomy, University of Birmingham,
Birmingham B15 2TT, UK
\newline
$^{  2}$Dipartimento di Fisica dell' Universit\`a di Bologna and INFN,
I-40126 Bologna, Italy
\newline
% ...
$^{  8}$CERN, European Organisation for Particle Physics,
CH-1211 Geneva 23, Switzerland
\newline
$^{  9}$Enrico Fermi Institute and Department of Physics,
University of Chicago, Chicago IL 60637, USA
\newline
% ...
$^{ 12}$Indiana University, Department of Physics,
Swain Hall West 117, Bloomington IN 47405, USA
\newline
% ...
$^{ 16}$Department of Physics, Schuster Laboratory, The University,
Manchester M13 9PL, UK
\newline
$^{ 17}$Department of Physics, University of Maryland,
College Park, MD 20742, USA
\newline
% ...
$^{ 23}$Department of Physics and Astronomy, Tel Aviv University,
Tel Aviv 69978, Israel
\newline
$^{ 24}$International Centre for Elementary Particle Physics and
Department of Physics, University of Tokyo, Tokyo 113-0033, and
Kobe University, Kobe 657-8501, Japan
\newline
% ...
$^{ 29}$University of British Columbia, Department of Physics,
Vancouver BC V6T 1Z1, Canada
\newline
% ...
 Debrecen, Hungary.
\newline

\section{Introduction} 
\label{sec:Intro}
The production of secondary heavy quarks from a virtual gluon is 
commonly referred to as gluon splitting. This process is 
considerably suppressed because
both the gluon and the quark jet from which it originates
must be sufficiently virtual to produce the heavy-quark pair. 
Nonetheless, these events make a significant contribution to 
heavy quark pair production in e+e- annihilation:
\eplemi\to\qqg, g\to${\mathrm Q\overline Q}$, where Q is a bottom or 
charm quark. These
events will be referred to here as g\to\ccbar\ or g\to\bbbar\ events.
This paper  describes a measurement of
the rate of g\to\ccbar\ at LEP at
center-of-mass energies in the region of the \Zz\ peak.

The probability of producing a heavy-quark pair from a gluon, per
hadronic Z$^0$ decay, is defined as
\begin{equation}
{\mathrm g_{Q\overline Q}}  = \frac{
  \mathrm{N}({\mathrm Z^0}\rightarrow \mathrm{q\overline qg,g}\rightarrow
{\mathrm Q\overline Q})}{ 
\mathrm{N(Z^0\rightarrow hadrons)}}.
\end{equation}
This probability has been calculated in the framework of perturbative QCD
to leading order in
$\alphas$, with the resummation of large leading and next-to-leading 
logarithmic terms to all orders~\cite{seymour,nason,seyi,miller}.
The probabilities for the secondary production of a charm-quark or bottom-quark
pair are predicted to be in the range
(1.35--2.01)$\times 10^{-2}$ for \gcc, and (1.75--2.90)$\times 10^{-3}$
for \gbb.
Precise measurements of these 
quantities allow an important comparison with QCD calculations, 
and also reduce the uncertainty in the experimental determination
of electroweak variables, such as
R$_{\mathrm c}$, the fraction of \Ztocc\ events in hadronic \Zz\ decays.

The first measurement of \gcc\ was made by OPAL~\cite{Dspm}, where
the contribution from the process g\to\ccbar\
to the inclusive \Dstar\ meson momentum spectrum was obtained by subtracting
the contribution from  \Ztocc\ and \Ztobb\ events 
where the \Dstar\ was produced from a primary quark.
This was followed by a second 
OPAL analysis~\cite{gcc}, in which gluon jets were 
selected, and charmed hadrons in these jets were identified by 
a lepton tag.
Combining the two 
measurements gave  \gcc$=$ (2.38\plm0.48)$\times 10^{-2}$~\cite{gcc},
a value which is compatible with the upper range of the theoretical predictions.

The ALEPH and DELPHI collaborations have recently measured the
rate of secondary production of b-quark pairs to be
\gbb=(2.77\plm 0.42\plm 0.57)$\times10^{-3}$~\cite{gbb1} and 
\gbb=(2.1\plm 1.1\plm 0.9)$\times10^{-3}$~\cite{gbb2}
respectively. Both
measurements are consistent with theoretical predictions.

The existing OPAL measurements~\cite{Dspm,gcc} only analysed
part of the total \Zz\ data sample. In this analysis the full sample
was used.
The precision of this measurement was further improved with a more
refined data calibration, optimisation of the analysis algorithms, and a
more reliable Monte Carlo simulation of the OPAL detector. 
To establish a signature for secondary charm-quark-pair production,
events containing three jets were selected, and the gluon jet identified.
The charm content of the gluon jet candidate was analysed by
identifying electrons, muons, and
\Dstar\ mesons, to give three independent measurements of g\to\ccbar.
The analysis is presented as follows: 
Section~\ref{sec:EvtSel} describes the hadronic sample and event simulation;
Section~\ref{sec:JetSel} discusses the selection of events likely to contain a
hard gluon and the 
methods used to select the gluon jet;
this is followed
by a description of the lepton and \Dstar\ channels separately 
in Sections~\ref{sec:LeptonAnalysis} and~\ref{sec:DstarAnalysis};
the systematic uncertainties for all the tagging schemes 
are listed in Section~\ref{sec:Sys}; the paper concludes with a
summary in Section~\ref{sec:summary}.

\section{Hadronic Event Selection and Simulation}
\label{sec:EvtSel}

We used data collected at LEP by the OPAL detector \cite{OPAL} between
1990 and 1995 in the vicinity of the \Zz\ peak.
Hadronic \Zz\ decays were selected using the number of charged tracks
and the visible energy in each event as in Reference \cite{evsel}. This
selection yielded 4.41 million events. 
The primary vertex of the 
event was reconstructed using the 
charged tracks in the event and the knowledge of the position and 
spread of the \eplemi\ collision point.

Monte Carlo events were used to determine the selection efficiency and 
background levels. 
The selection efficiency was measured in dedicated samples of events
containing the g\to\ccbar\ process. 
For the lepton analysis, 
at least one of the charmed hadrons 
was required to decay semileptonically. The
\Dstar\ analysis used events where at least one of the secondary charm quarks
hadronised to a \Dstar\ which decayed via ${\mathrm D^{\star +}}$\to\Dz\Pip, 
followed
by \Dz\to\Km\Pip\footnote{Charge conjugation is assumed throughout this paper.}.
For background
studies, 9 million 5-flavour hadronic \Zz\ decays plus an additional
3.5 million \Ztobb\ events and 2.5 million \Ztocc\ events were
generated. All these samples were produced with
the JETSET 7.4 Monte Carlo program \cite{seventhree}. The heavy quark
fragmentation was parametrized by the fragmentation
function of Peterson
\etal\ \cite{peterson}, and
the measured values of the
partial widths of \Zz\ into \qqbar\ were used \cite{LEPSLD}. The production
rates of different charmed hadrons at $\sqrt{s}=$ 10\thinspace\GeV\ 
and $\sqrt{s}=$ 91\thinspace\GeV\ are consistent~\cite{pdg}, so
the mixture of charmed hadrons 
produced in \Ztocc, g\to\ccbar ,  and in b hadron decays were
taken from Reference~\cite{pdg}, as were
the semileptonic branching ratios of charm 
hadrons. All samples were processed with the 
OPAL detector simulation package~\cite{gopal}.

\section{Jet Selection}
\label{sec:JetSel}
\subsection{Jet Reconstruction}
Measurement of secondary charm pair production could in principle be 
reduced to a charm
hadron counting experiment in events where the primary quarks
were light flavoured, i.e.\ up, down, or strange quarks. However,
such a sample is hard to obtain as 
\Ztocc\ and \Ztobb\ events can not be identified with 100\thinspace\%
efficiency. Instead, by grouping the particles of an event into jets,
charmed hadrons which are produced from the primary quarks can more
easily be identified and
counted as background. 

To identify the process of gluon splitting into a charm-quark pair, we
required the event to contain exactly three jets.
Using simulated events, we investigated the effect of various
jet finding algorithms to select the three-jet topology. Of the 
Durham, Geneva, Cone, and JADE-E0 algorithms~\cite{otherjets} that
were tested, the JADE-E0
recombination scheme with a $y_{\mathrm {cut}}$ value of
0.05 gave
the most significant g\to\ccbar\ signal.
As can be seen from Figure \ref{fig:jets}(a), the two secondary charm
quarks tend to be contained within a single jet, so that in three-jet
events, g\to\ccbar\ events were
identified more efficiently than background events.

For the selected three-jet events, the jet energies were calculated,
neglecting mass effects, using the relation
\begin{equation}
E_i=E_{\mathrm {cm}}{\sin\psi_{jk}\over
  \sin\psi_{jk}+
\sin\psi_{ij}+\sin\psi_{ik}},
\end{equation}
where $E_{\mathrm {cm}}$ is the center-of-mass energy,
$\psi_{ij}$ is the angle
between jets $i$ and $j$ and $E_i$ is the calculated energy 
of jet $i$. This equation only holds for 
coplanar events and therefore,
if the sum of the angles between the jets was smaller
than $358^\circ$, the event was rejected.
These criteria retain 30\thinspace\% of the data sample. 
From the simulation, the efficiency for
g\to\ccbar\ events to pass the selection is 56\thinspace\%.

In the previous OPAL analysis \cite{gcc},
only 3.5 million hadronic \Zz\ decays were used, with a 
$y_{\mathrm {cut}}$ value of 0.03, compared to 4.4 million events
and a $y_{\mathrm {cut}}$ value of 0.05 used here.

\subsection{Gluon Jet Selection}
The gluon jet candidate in the event was then selected. Several
selection algorithms were tried, of which two were retained, 
based on their efficiency for selecting the correct jet and
for background rejection.
The first method, already used in Reference~\cite{gcc}, assumes that
the lowest energy (LE) jet is the gluon jet. 
The second method, used here for the first time, takes the  
jet that is most readily subdivided into two as the 
gluon jet candidate (JS), and was found
to correctly identify the gluon jet in g\to\ccbar\ events more often.
This method is motivated by the fact that the gluon jets of interest
contain two charmed hadrons, and should therefore show more evidence
of jet substructure than 
primary charm jets and most bottom jets which contain only one
charmed hadron.
The JADE-E0 jet finding algorithm was applied to each jet individually,
and the gluon jet candidate is taken to be the one with the highest
value of $y_{\mathrm cut}$ at which the jet splits into two subjets.
Using the new JS method, 
the correct gluon jet is identified in 60\thinspace\% of three-jet events from
the signal process g\to\ccbar, while the 
corresponding number for the LE method is 51\thinspace\%.

The performance of the two gluon selection schemes was investigated for
both the lepton and the \Dstar\ analyses. The LE method performed better
than the JS method for the lepton analysis, after all the 
additional background rejection criteria had been applied (section 4.2).
The overall efficiency of the lepton analysis using the LE method
was 66\thinspace\% of the efficiency using the JS method, but with a 
factor of three less background from primary heavy-quark events, yielding 
a 15\% advantage in statistical significance.
In contrast, for the \Dstar\ analysis, the JS method outperformed the LE
method; the efficiency for the \Dstar\ analysis doubled  when using the
JS method while the heavy quark background only increased by a quarter, 
yielding an improvement in statistical significance of 80\%.
As a results of these studies, the LE method was used for the lepton analysis
while the JS method was used for the \Dstar\ analysis.

\begin{figure}[ht]
   \begin{center}\hspace*{-.4cm}\mbox{
        \epsfig{file=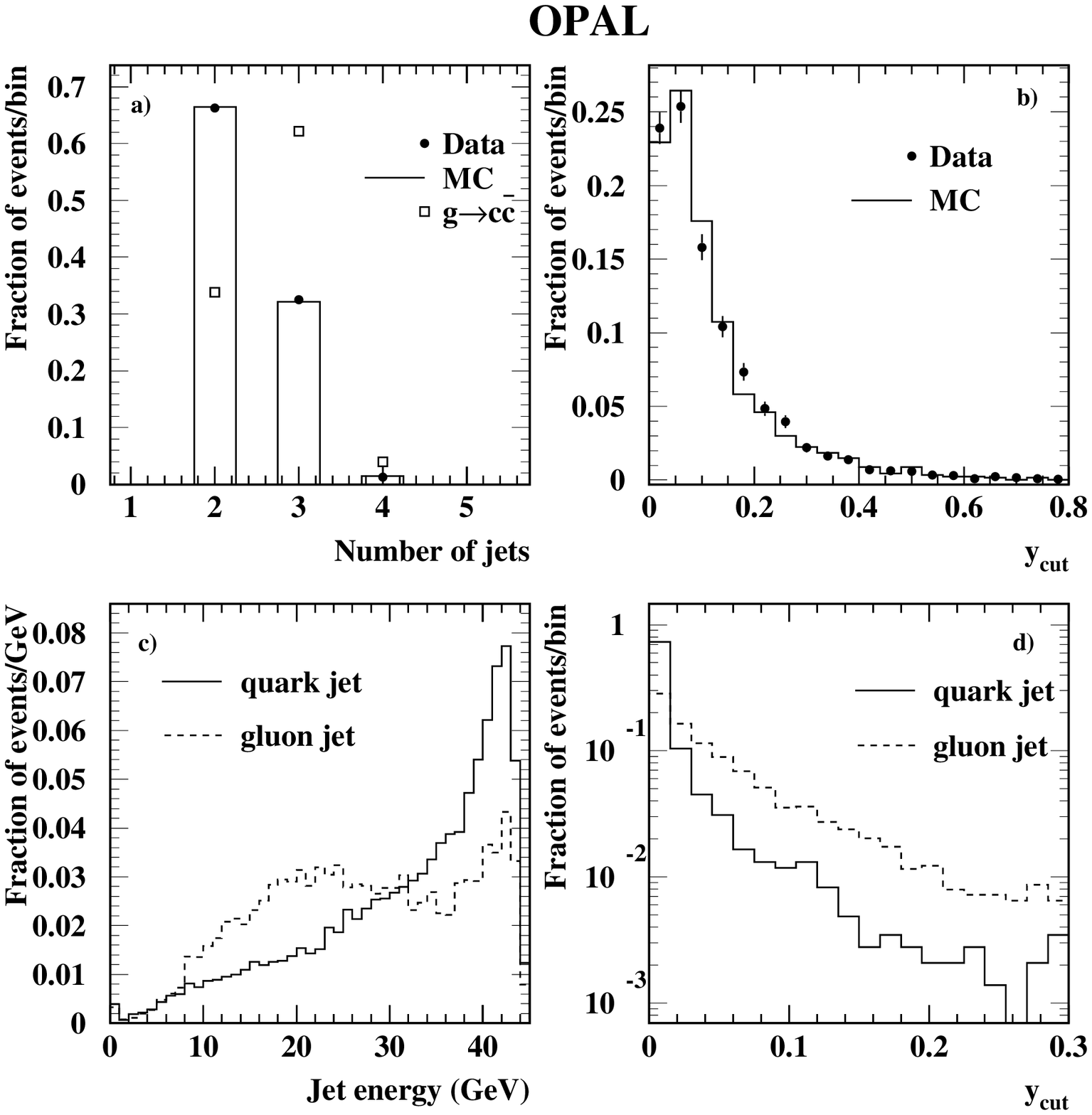,height=18cm}}
   \caption{Jet selection properties: a) number of jets in the events;
     b) $y_{\mathrm cut}$ value where a selected jet splits into two 
     subjets for all three-jet events; c) jet
     energy for quark and gluon jets in Monte Carlo g\to\ccbar events;
     d)  $y_{\mathrm cut}$ value at
     the point where a jet splits into two subjets for quark and gluon
     jets.
     The Monte Carlo histograms in a) and b) include the default JETSET
     value of \gcc=0.014.}
   \label{fig:jets}
\end{center}
\end{figure}

Figure \ref{fig:jets}(b) shows the  $y_{\mathrm cut}$ value of the 
selected jet at the point where that jet splits into two subjets, for 
all three-jet events selected from data and the Monte Carlo simulation.
Figure \ref{fig:jets}(c) shows the distribution of the jet energy for quark and
gluon jets from the process g\to\ccbar, and  Figure \ref{fig:jets}(d) 
shows
the  $y_{\mathrm cut}$ value at the jet splitting point for these jets.

\section{The Lepton Analysis}
\label{sec:LeptonAnalysis}
Using the sample of three-jet events, and taking the gluon jet candidate 
to be the lowest energy jet (LE method), events containing the gluon
splitting process, g\to\ccbar, were searched for 
using a lepton tag, which assumes that one 
of the charm quarks decayed semi-leptonically.

\subsection{Lepton Identification}
\label{sec:LeptonAnalysis:id}
Electrons were identified using an artificial neural network~\cite{Rbmain}
while muons were identified using 
information from the muon chambers in association with the tracking chambers, as in~\cite{muon}. 
The lepton identification was limited in polar angle $\theta$, defined
as the  angle between the lepton candidate and the beam axis direction 
of outward-going electrons. The polar angle of identified electrons  
was limited to $\vert\cos\theta\vert<0.8$, while identified muons were
required to be
within  $\vert\cos\theta\vert<0.9$. In addition, the momentum of the lepton candidate was
 required to satisfy  $ 2<p<9$\thinspace\GeVc, and have a transverse momentum, 
with respect to the jet axis containing the candidate, below 2.75\thinspace\GeVc . 
Electrons consistent with being produced from photon conversions or from Dalitz decays
were rejected with an artificial 
neural network using geometrical and kinematical properties~\cite{Rbmain}.

The efficiency for leptons from the g\to\ccbar\ signal
events to pass the lepton identification and the momentum
selection criteria was 16\thinspace\% for electrons, and
17\thinspace\% for muons. 
This differs from the 
previous OPAL analysis~\cite{gcc}, with the most significant
difference being the larger momentum range used. Previously
muons were only accepted in the narrower range of $3<p<6$\thinspace\GeVc. 
The shift to the larger momentum range, together with the improvements
in electron identification and the photon
conversion finder, have resulted in a larger efficiency for
identifying leptons from g\to\ccbar.

At this point, the data sample contained $7\,180$ tagged electron
candidates and
$14\,631$ tagged muon candidates, corresponding  to a \gcc\  purity of
10\thinspace\% and 6\thinspace\% 
 respectively, while the efficiency to detect a lepton from the
g\to\ccbar\ process that passes all the criteria described above (Sections
2-4.1) was 
approximately 7.5\thinspace\%. The large difference between the
number of selected
electrons and muons is mostly due to contamination of hadrons passing the muon 
selection criteria, as described in section 4.3.

\subsection{Suppression of Jet Mis-assignment Background}
\label{sec:LeptonAnalysis:supp}
With the selection of lepton tagged events described above,
background contributions from events with jet mis-assignment  
were unavoidable: Jet mis-assignment events are defined as those where the 
lowest energy jet contains a lepton from the decay of a primary heavy quark 
rather than from a heavy quark in a gluon jet.
The jet mis-assignment background is dominated by \Ztobb\ events, and was
suppressed by means of a b-tagging algorithm based on reconstructed
displaced secondary vertices. A neural network with inputs based on
decay length significance, vertex multiplicity and invariant mass 
information \cite{VNN} was used to select
vertices with a high probability of coming from b hadron
decays. Events were rejected if any of the three
jets were tagged by the neural network.
This procedure resulted in a reduction of the jet
mis-assignment background by 43\thinspace\%, which included a 54\thinspace\%
reduction in \bbbar\ events, while retaining 78\thinspace\% of the g\to\ccbar\ 
sample. After this cut the data sample contained 5\,049 electron candidates and
11\,031 muon candidates.

To further reduce the jet mis-assignment background from non-gluon jets, we explored
several properties of the lowest energy jet, with respect to the other
two. Naively, as a jet containing the  gluon splitting g\to\ccbar\ contains 
two charm quarks, one would expect the jet mass and multiplicity to be larger  
for that jet than for the other two jets.
Comparison of Monte Carlo samples containing the signal, with samples 
of \Ztobb\ and \Ztocc\ events in which 
a lepton coming from the decay of a primary b or c quark
was assigned by the jet finder to the lowest energy jet showed that background 
rejection through mass and multiplicity cuts was indeed feasible, and the 
following requirements were found to give the best
background rejection:
\begin{itemize}
\item ${{\mathrm Max (M_1,M_2)} / {\mathrm M_3} < 2 } $, where  M$_i$
is the mass of the $i$th jet, and the jets are ordered by energy,
with jet 1 having the highest energy.
\item $ {{\mathrm (N_1 + N_2)} / {\mathrm N_3} < 2.5}$, where N$_i$ 
is the
track and electromagnetic cluster multiplicity of the
${i}$th jet, where an electromagnetic cluster was counted only if no
charged track was associated with it. 
\end{itemize}

These two selection criteria retained 55\thinspace\% of the g\to\ccbar\ events,
while rejecting 71\thinspace\% of the \Ztocc\ background and
67\thinspace\% of the \Ztobb\ background.
On application of the jet mis-assignment background suppression, in
conjunction with the lepton tagging selection criteria described previously,  
and accepting only one
lepton tag candidate in each event (with priority given to electrons
due to the
higher muon fake rate), the data sample was reduced to $2\,434$
electron candidate events and $4\,362$ muon candidate events.
At this stage, no further background suppression was  done. Rather,
the remaining backgrounds in the data sample were evaluated using
Monte Carlo estimates (see Section 4.3).

\subsection{Estimates of Background Rates}

The determination of the rate of each background source is 
discussed below. 
A summary of the estimated data sample composition is given in Table 1.

\subsubsection*{Jet Mis-assignment Background}
The rate of the jet mis-assignment background was estimated from 
the 14.5 million hadronic \Zz\ decays of the Monte Carlo 
sample mentioned in Section 2,
after applying all the selection cuts of Section 4.2.
The number of leptons 
from semileptonic decays of charm and
bottom hadrons was determined, excluding leptons from 
g\to\ccbar\ or g\to\bbbar\ decays. By 
scaling to the number of hadronic events in the data sample, we estimated
this background to contain $1\,027$\plm 32 candidates from \Ztobb\ events 
and 702\plm26 candidates from \Ztocc\ 
events, where these uncertainties are statistical only.

\subsubsection*{Photon Conversions}
From the Monte Carlo simulation, the photon conversion finder fails
to tag (14.3\plm 0.4)\thinspace\% of the conversions, where the
uncertainty is from the systematic uncertainties in the conversion finding
efficiency~\cite{Rbmain}. These untagged conversion electrons then form a 
background to the electron tagged events.
Knowing the efficiency of the conversion finder, the background
from untagged conversions was estimated from the number of tagged
conversions to be 
630\plm25 events.

\subsubsection*{Lepton Mis-identification and Decays in Flight}
To estimate the background from hadrons which were erroneously 
identified
as lepton candidates, we used the Monte Carlo sample to determine the
probability that
a charged track with a given momentum, $p$, and transverse momentum with 
respect to the 
direction of the associated jet, $p_t$, should be incorrectly identified
as a lepton. The number of background leptons in the data was then derived
by multiplying the number of tracks in the data that passed the
selection criteria, excluding the lepton identification, by these fake
probabilities. 
In practice the fake probabilities per track are estimated in bins of $p$ and
$p_t$, and corrected  for differences between the Monte Carlo simulation and 
the data, as in \cite{muon}. 
The correction for the difference between the MC and the data introduces
a large uncertainty on the number of hadrons mis-identified as leptons as 
described in section 6.1

The total number of hadrons mis-identified as muons was estimated at 
$2\,580$\plm51.
Decays in flight of light hadrons into muons are included in this 
estimate. The number of hadrons
mis-identified as electrons was estimated by a similar method
to be 81\plm9.


\subsubsection*{Dalitz Decays of ${\mathbf \pi}^{\mathbf 0}$ and
    $\mathbf \eta$}
The number of background events from the decay of $\pi^0$ and $\eta$ into 
\eplemi$\gamma$ was estimated from Monte Carlo simulation and
corrected to the known
$\pi^0$ and $\eta$ multiplicities in \Zz\ decays \cite{pdg}. These 
contributions were estimated  at 149\plm12 electron candidate
events.

\subsubsection*{Gluon Splitting: $\mathrm{\bf g\rightarrow b\overline{b}}$}
The number of events from the process g\to\bbbar\  that survive the selection 
criteria was calculated from
$ \gbbd \cdot{\mathrm N_{\mathrm had}} \cdot \epsilon^{\mathrm b} $, 
where $\epsilon^{\mathrm b}$ is the efficiency for at least one lepton  from 
the process g\to\bbbar\ to survive the selection criteria,
 ${\mathrm N}_{\mathrm had}$   is the number of 
hadronic events, and  $\gbbd = (2.69 \pm 0.67)\times 10^{-3}$ 
is the averaged measured
value of  \gbb\ taken from  \cite{LEPSLD}.
From Monte Carlo simulation, $\epsilon^{\mathrm b}$ was found to be
$(1.0\pm0.1)\thinspace\%$, leading to an estimated background of 
103\plm10.

\begin{table}
\begin{center}
\begin{tabular}{|l|c|c|} \hline
Quantity & Electron channel & Muon channel \\
\hline
Observed events & 2\,434 & 4\,362 \\
\hline
Jet mis-assignment (\ccbar) & 342\plm19  & 360\plm18 \\
Jet mis-assignment (\bbbar) & 494\plm22  & 533\plm23 \\
Residual photon conversion & 630\plm25 & - \\
Lepton mis-identification & 81\plm9 & 2\,580\plm51 \\
Dalitz decays & 149\plm12& - \\
${\mathrm g\rightarrow b\overline b}$ & 50\plm7 & 53\plm7 \\
\hline
Estimated signal & 688\plm42 & 836\plm59 \\
\hline
\end{tabular}
\caption{Summary of selected sample sizes and estimated composition.}
\label{tab:result}
\end{center}
\end{table}

\subsection{Comparison of the Data and Monte Carlo}
\label{sec:LeptonAnalysis:datamc}
Since a major part of the background contribution has been estimated from the Monte
Carlo, it was crucial that the consistency between the 
Monte Carlo and the OPAL data be verified.
Of particular importance was the accuracy in the simulation and estimation  of the jet
mis-assignment background, as the efficiency to tag a heavy quark jet as a
g\to\ccbar\, and the fraction of heavy quark jets found in the lowest
energy jet were critical to this analysis.

To check the validity of the jet mis-assignment background composition, 
we examined the
lepton yield in three-jet events where the lepton did not originate
from the lowest energy jet. 
Specifically, we searched for events with leptons in either of the two highest 
energy jets which passed  all but the lepton requirement in the third jet of the 
selection criteria. The leptons were identified
using the same 
criteria as described in Section \ref{sec:LeptonAnalysis:id}, and  the jet-based mass 
and multiplicity cuts were applied  in the same fashion as in Section  
\ref{sec:LeptonAnalysis:supp}. 
However, the search was restricted to prompt leptons from b hadron
decay by requiring lepton momenta above 5 
\GeVc\ and transverse momenta above 1.5 \GeVc.
The resulting yield of 
such leptons per three-jet event was found to be $(1.49\pm0.02)\times 10^{-3}$ 
in the Monte Carlo and 
$(1.47\pm0.02)\times10^{-3}$ in the data. For the
Monte Carlo sample the simulation showed that the b-hadron purity of
these events was 85\thinspace\%. 

We also looked at events tagged as \Ztobb\ with the b-tagging algorithm
described in Section~\ref{sec:LeptonAnalysis:supp} in order to compare 
the jet mis-assignment rate in
\Ztobb\ events. This was done by searching for a lepton in the lowest
energy jet of events that have vertices compatible with b-hadron decay. 
After applying all other selection criteria
we observed 646 such events in the data. The Monte Carlo prediction
scaled to the data sample size was found to be 632 events. The  Monte Carlo
simulation estimated that the b hadron purity of these events 
was approximately 93\thinspace\%. From  the agreement of the Monte Carlo prediction with the 
data sample estimates for these two tests, the data/Monte Carlo consistency was found
to be adequate, and the jet mis-assignment  well modelled in the 
Monte Carlo.

\begin{figure}[ht]
   \begin{center}\hspace*{-.4cm}\mbox{
        \epsfig{file=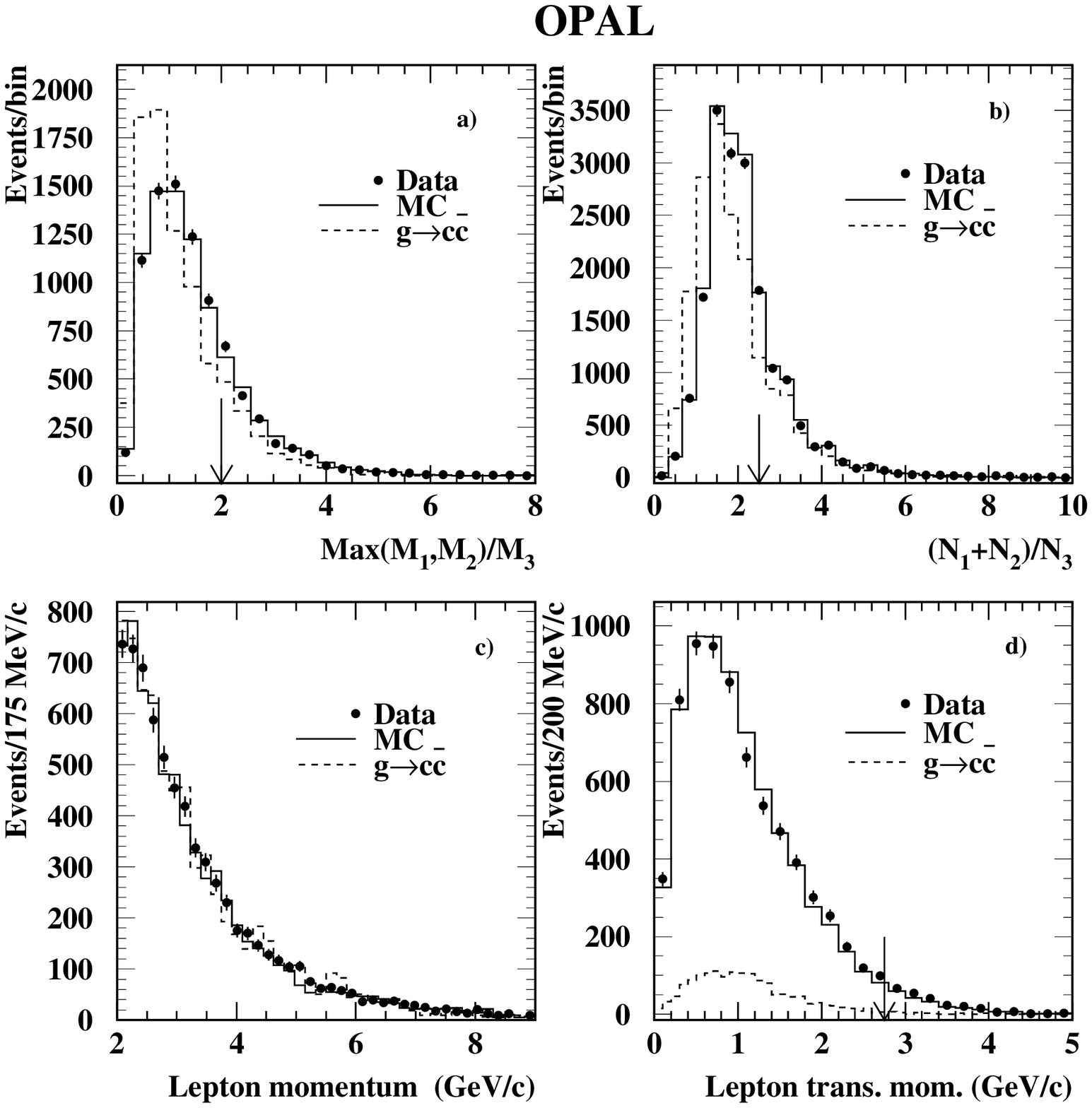,height=18cm}}
   \caption{Comparison of data and Monte Carlo for the lepton analysis
(histograms represent the 
Monte Carlo and data are represented by the points, the dashed
histograms show the g\to\ccbar\ spectrum). 
Monte Carlo distributions (solid and dashed histograms) are 
normalised to the number of hadronic events in the data. 
a) Ratio of the maximum mass of the first two jets with respect to the third
jet for events that passed the lepton selection and the b-tagging
neural network rejection;
b) ratio of the multiplicity of tracks and unassociated clusters in
the first two jets with respect to the third jet for the events in a);
c) lepton candidate spectrum for events passing all the selection criteria;
d) lepton candidate transverse momentum spectrum for events passing all
but the transverse momentum criteria with the g\to\ccbar\ spectrum
scaled to the signal area. The arrows in a), b) and d) show the cut value
used in this analysis. Events below these values were accepted.
}
   \label{fig:mcdata}
\end{center}
\end{figure}

In addition,  comparisons of the Monte Carlo simulation prediction and
data distributions for the variables critical to the lepton tagging
analysis were made, with the results of the comparison shown in
Figure~\ref{fig:mcdata}. Specifically, Figure~\ref{fig:mcdata}(a) shows 
the ratio of the maximum mass of the first two jets with respect to the third
jet. Figure~\ref{fig:mcdata}(b) 
shows the ratio of the multiplicity of tracks and unassociated clusters in
the first two jets with respect to the third jet. Figure \ref{fig:mcdata}(c) 
shows the lepton candidate momentum spectrum for events passing all the 
selection criteria, and \ref{fig:mcdata}(d) shows
the lepton candidate transverse momentum.
All plots present the Monte Carlo prediction
which include the g\to\ccbar\ contribution normalised to the number of
signal events obtained in this analysis.
This comparison shows an agreement between Monte Carlo and data in
both shape and rate prediction. Thus, the procedure of subtracting the
Monte Carlo prediction for the jet mis-assignment background was
justified.

\subsection{Results}
\label{sec:results}

\begin{figure}[ht]
   \begin{center}\hspace*{-.4cm}\mbox{
        \epsfig{file=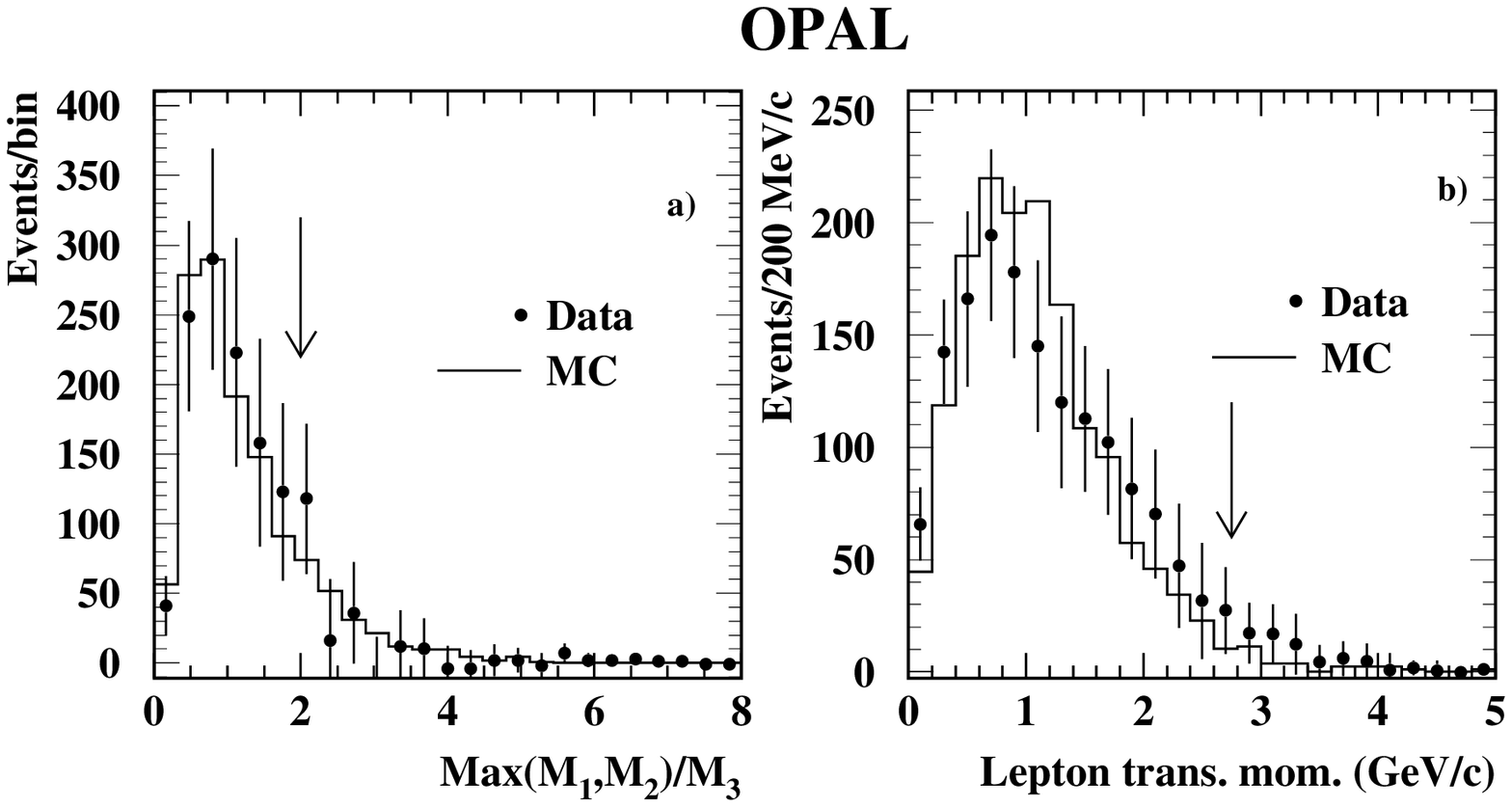,height=10.5cm}}
   \caption{Distributions of  a) ${\mathrm Max(M_1,M_2)/M_3}$, 
     b) lepton transverse momentum for background subtracted data (points) and
     for g\to\ccbar\ Monte Carlo scaled to the average lepton analysis
     \gcc\ result (solid line):
     All the selection criteria described in the text were applied with the
      exception of the variable in each plot where the cut is
     shown by the arrow. The error bars shown represent the
     statistical and systematical uncertainties combined.
}
   \label{fig:signal}
\end{center}
\end{figure}

The charm-quark-pair production rate per hadronic event is related to the 
measured quantities by
\begin{equation}
\mathrm g_{c\overline c}  = {{\mathrm N}_{\mathrm {sel}}
\over {\mathrm N}_{\mathrm {had}}\cdot\epsilon\cdot2\cdot 
{\mathrm B}({\mathrm c}\rightarrow {\mathrm {X\ell}}\nu)},
\end{equation}
 where the following notation is used; ${\mathrm N}_{\mathrm {sel}}$ 
is the number of events passing the 
selection criteria after subtraction of background events (Table 1), ${
\mathrm N}_{\mathrm {had}}$ is the 
number of hadronic \Zz\ decays,  $\epsilon$ is the
efficiency for finding a single lepton from a sample of g\to\ccbar\ Monte Carlo
events in which at least one of the charmed
hadrons decayed semileptonically and passed the selection criteria, and
${\mathrm B}({\mathrm c}\rightarrow {
\mathrm {X\ell}}\nu)$ 
is the charm hadron semileptonic branching 
ratio of $(9.5\pm0.7)\thinspace\%$ obtained by taking the average of
the most recent measurements of
OPAL\cite{OPALcsl} and
ARGUS\cite{ARGUS}.
With ${\mathrm N}^{\mathrm e}_{\mathrm {sel}}=688\pm42$, 
${\mathrm N}^{\mu}_{\mathrm {sel}}=836\pm59$,
$\epsilon^{\mathrm e} = (2.72\pm0.12)\thinspace\%$ and $\epsilon^{\mu} = 
(2.83\pm0.13)\thinspace\%$  we obtained
\begin{eqnarray}
{\mathrm g^{\mathrm e}_{c\overline c}}& = 0.0309\pm 0.0029, \\
{\mathrm g^{\mu}_{c\overline c}}& = 0.0360\pm 0.0038, 
\end{eqnarray}
where the  uncertainties are the statistical contributions only.
Figure \ref{fig:signal}(a) gives the comparison between the 
background subtracted data
distribution of the three-jet mass variable, and the JETSET prediction
for the same 
distribution (normalised to the data sample). A clear enhancement in the
mass  distribution is visible at low values of the mass ratio, which
is well described by the Monte Carlo prediction, and when compared to 
Figure \ref{fig:mcdata}(a), justifies the cut used in section 4.2. 
Furthermore, the
agreement between data and Monte Carlo suggests that JETSET describes the
production and shape of secondary heavy quarks rather well.
Figure \ref{fig:signal}(b) shows the spectrum of the lepton transverse
momentum for the background subtracted data and for the g\to\ccbar\ 
Monte Carlo
simulation scaled to the data size. Here too, a reasonable agreement
between the  data and Monte Carlo is seen.
Leptons were accepted if their transverse momentum was
smaller than 2.75 \GeVc\ (where the majority of the signal was found). 

\section {The ${\mathrm D^{\star\pm}}$ Analysis}
\label{sec:DstarAnalysis}
As \Dstar\ mesons are copiously produced from charm quarks and have a clear
signature, they can be used to tag charmed hadrons in a gluon jet.
Using the three-jet event sample, we chose the gluon jet
candidate using the jet splitting technique described in section 3.2,
and searched for a \Dstar\ meson in this jet.
The D$^{\star +}$\to D$^0$\Pip, D$^0$ \to \Km \Pip\  decay channel
gives a clean signal, since 
the small mass difference between the \Dstar\ and the
\Dz\ limits the phase space available, reducing combinatorial background.

The \Dz\ reconstruction was performed as in \cite{Dspm} by trying all 
combinations of
oppositely charged tracks, assuming one of them to be the kaon, and the
other to be the pion. We then added a third track, the  ``slow pion'', 
demanding its charge
to be equal to that of the pion candidate track, to
form the \Dstar\ candidate.

The following mass cuts were applied:
\begin{itemize}
\item The reconstructed D$^0$ mass must lie within 75\thinspace\MeV\
   of the nominal
   D$^0$ mass.
\item The mass difference, $\Delta M$, between the \Dstar\
  and the \Dz\ candidates must be in the range \\
$0.143\thinspace\GeVcc~<~ \Delta M~ <~ 0.147~ \GeVcc.$ 
\end{itemize}
To further reduce the background from random combinations, 
the following criteria were imposed:
\begin{itemize}
\item The measured rate of energy loss for the kaon candidate track
was required to be consistent with that expected for a kaon with
a probability of more than 0.1.
\item The kaon candidate track momentum satisfied $p_{\mathrm K} >1.5$\thinspace\GeVc.
\item The ratio between the
measured \Dstar\ energy to the total energy measured in the selected 
jet satisfied
${\mathrm {X_{jet}} ={{E_{\PDst}} \over {E_{jet}}}} > 0.2$.
\item The helicity angle, $\th^*$, measured between the kaon in the
D$^0$ rest frame and the D$^0$ direction in the laboratory frame
satisfied $\costhe^* < 0.7$. The
D$^0$ decays isotropically in its rest frame creating a flat
distribution of $\costhe^*$, while background events are peaked at 
$ \costhe^*=1 $. 
\end{itemize}
Charmed hadrons which pass these selection criteria can also be
produced in b hadrons decays. 
The artificial neural network described in Section \ref{sec:LeptonAnalysis:supp}
was used to reject such events.

\begin{figure}[ht]
   \begin{center}\hspace*{-.4cm}\mbox{
        \epsfig{file=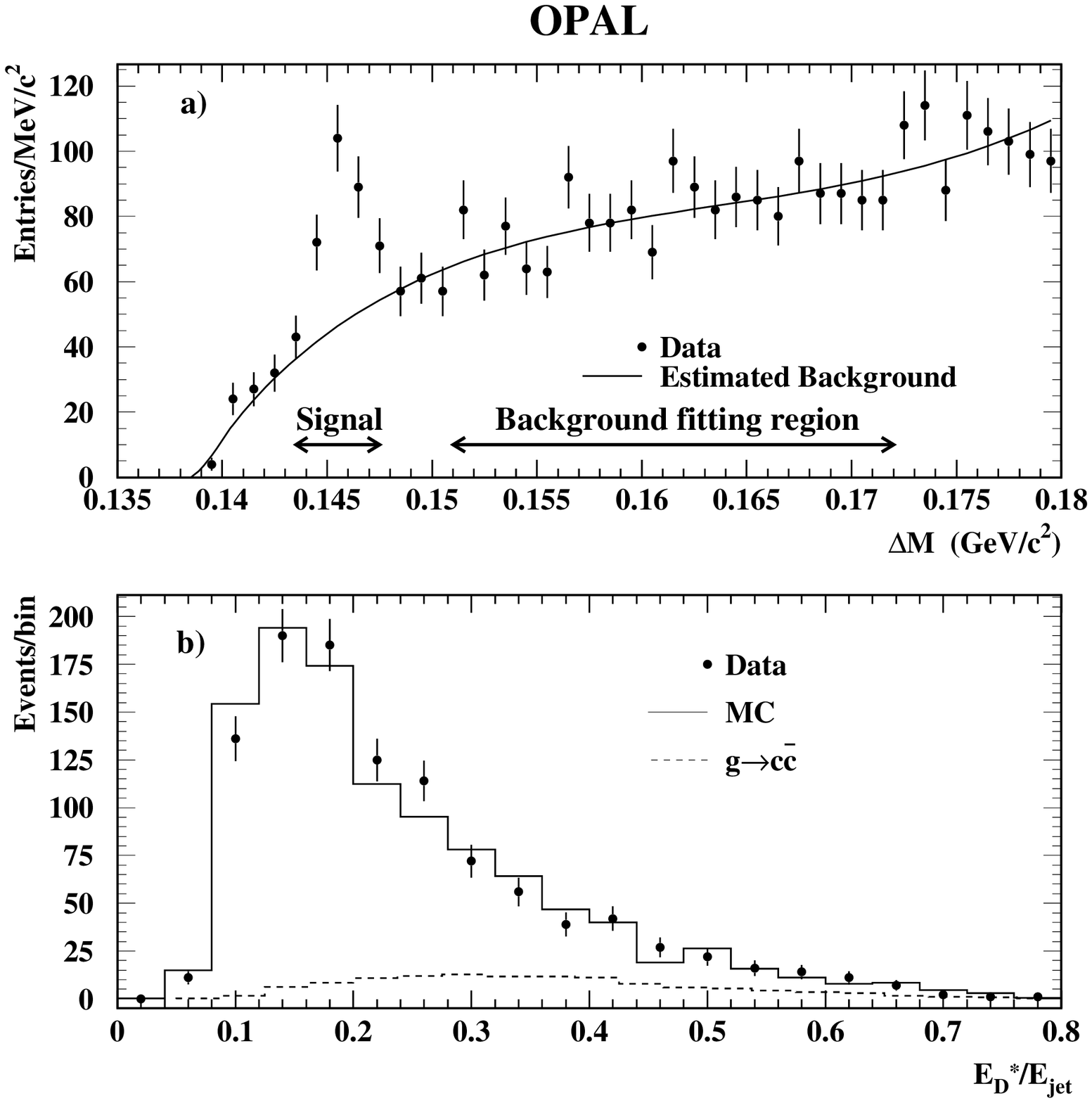,height=18cm}}
    \caption{ (a) The distribution of the difference of the invariant mass
     between the \Dstar\ candidate and the \Dz\ candidate, 
$\Delta M$. The solid curve 
represents the estimated
background shape. An estimated 137\plm 13 \Dstar\ candidates are above the 
background fit around 0.145 \GeVcc. 
(b) Ratio of the \Dstar\ candidate energy to the jet energy for
 candidates in the selected gluon jet after applying the combinatorial
 rejection criteria. The
 solid line shows the Monte Carlo spectrum with the g\to\ccbar\
 component scaled to measured value in the \Dstar\ analysis (\gcc=0.04), and
 the dashed
 histogram shows the spectrum for true \Dstar\ mesons from g\to\ccbar\
 scaled to the measurement area. }
   \label{fig:dstmass}
\end{center}
\end{figure}

Applying all
these selection criteria reduced the three-jet data sample from 1.32
million events to 308 events. This large reduction in sample size is
partially due to the low branching ratio of the decay chain as well as
the result of the background suppression criteria.
The shape of the remaining combinatorial background was described
by a function of the form
$(\Delta M-0.139)^a(b+c\Delta M+d(\Delta 
M)^2),$ 
with $a$, $b$, $c$, and $d$ determined from the sideband of the \Dz\ mass 
distribution with the signal being found in the 2.16~\GeVcc 
$<~{\mathrm M_{K^-\pi^+}}~<$ 2.46~\GeVcc\ region.
The normalisation of the background was determined outside the signal
region. The estimated background in the signal region 
was estimated to be 171.1\plm 13.1 events.
The distribution of $\Delta M$ is shown in 
Figure~\ref{fig:dstmass}(a) where a significant signal above the fitted
background estimation is observed around 145 \MeVcc.

The jet mis-assignment background was estimated from the Monte Carlo
simulation as in 4.3. We estimated this background to be 35.9\plm6.0 and
19.2\plm4.3 events from \ccbar\ and \bbbar\ events respectively. 
 
The remaining source of background considered was from 
secondary b quark pair
production (g\to\bbbar). To estimate its contribution, an average value of 
 $\gbbd = (2.69 \pm 0.67)\times
10^{-3}$~\cite{LEPSLD} was used, along with a selection efficiency of
0.034\thinspace\% (including branching ratios),
resulting in a g\to\bbbar\  background estimate of 4.0\plm2.0 events.
A summary of the sample composition is given in Table~\ref{tab:resulti}.

\begin{table}
\begin{center}
\begin{tabular}{|l|c|} \hline
Observed events & 308 \\
\hline
Combinatorial background  & 171.1\plm13.1  \\
Jet mis-assignment  (\bbbar) & 19.2\plm4.3 \\
Jet mis-assignment  (\ccbar) & 35.9\plm6.0 \\
${\mathrm g\rightarrow b\overline b}$ & 4.0\plm2.0 \\
\hline
Estimated signal & 77.8\plm15.2 \\
\hline
\end{tabular}
\caption{Summary of observed events and estimated background for the
  \Dstar\ analysis. }
\label{tab:resulti}
\end{center}
\end{table}

In a similar fashion to Section \ref{sec:LeptonAnalysis:datamc}, 
we have compared the Monte Carlo
simulation with the data. Figure \ref{fig:dstmass}(b) shows the 
 ratio of the \Dstar\ candidate energy to the jet energy for
 candidates in the selected gluon jet after applying the combinatorial
 rejection criteria. The Monte Carlo distribution contains a
 g\to\ccbar\ component scaled to the result obtained for this channel 
(\gcc\ = 0.04). 

To obtain the rate of g\to\ccbar\ we modified equation 3 to account
for the c\to \Dstar , \Dstar \to
\Dz\Pip and \Dz\to $\mathrm{ K}^-\pi^+$ branching ratios.
The product of
the first two was taken to be (15.27\plm0.92)\thinspace\%~\cite{LEPSLD}  and 
B(\Dz\to$\mathrm{ K}^-\pi^{+})$ to be (3.85\plm0.09)\thinspace\%~\cite{pdg}. The number
of selected events was
${\mathrm N}_{\mathrm {sel}}=77.8\pm15.2 $, and the selection  efficiency
found from the Monte Carlo was
$\epsilon =(3.70\pm0.12)$\thinspace\%. This gives
\begin{equation}
{\mathrm g_{c\overline c}} = 0.0408\pm 0.0122,
\end{equation}
where the uncertainty is statistical.

\section{Systematic Uncertainties} 
\label{sec:Sys}
Possible sources of systematic uncertainty and their effect on the
background and efficiency are discussed below. 
The systematic uncertainties on \gcc\ from the different sources are 
summarized in 
Table~\ref{tab:sys} separately for the electron, muon and \Dstar\ measurements.

\subsection{Lepton Identification}
\label{sec:SysLepton}

{\bf Muon identification efficiency:} The systematic uncertainty from the 
muon identification efficiency was evaluated using a 
method similar to that in Reference~\cite{muon}.  
The muon identification efficiency was compared in data and Monte
Carlo using various control samples, including \Zz\to$\mu^+\mu^-$
events, and muons reconstructed in jets.
Without using $\dEdx$ information, 
an uncertainty of $\pm 2.0\thinspace\%$ was found. 

However, as $\dEdx$ information is an important input to muon
identification,
the effect of mis-modelling of the $\dEdx$ was studied.
The mean $\dEdx$ for muons 
in \Zz\to$\mu^+\mu^-$ events was observed to be shifted by 
approximately $15\thinspace\%$ of the $\dEdx$ resolution with respect to
the theoretically expected value. A similar shift was observed 
in the Monte Carlo simulation, both for muon pairs and for muons 
identified inclusively in jets, and an uncertainty on the modelling of
the $\dEdx$ mean of
$\pm5\thinspace\%$ was assigned.
The $\dEdx$ resolution was studied in the data and Monte Carlo using
test samples and was found to be modelled to better than $5\thinspace\%$. 
The uncertainties on these sources gave a total uncertainty on
the efficiency of muon identification of $\pm 3.0\thinspace\%$.\\

\noindent{\bf Muon mis-identification rate:} There was an uncertainty on the
number of hadrons that were incorrectly identified as muons.
This predominantly arises from uncertainties in the simulation of the 
fake probabilities. Three samples were used to test the Monte Carlo
modelling of these probabilities:
\begin{itemize}
\item pions from identified $\PKs$\to$\pi^+\pi^-$ decays. 
The Monte Carlo predicted that the tracks in this sample were more than 99\thinspace\% 
charged pions with less than 0.1\thinspace\% 
contamination from prompt muons.
\item three-prong $\tau$ decays. 
The Monte Carlo predicted that the tracks in this sample were more than 99\thinspace\% 
charged pions with less than 0.1\thinspace\% 
contamination from prompt muons.
\item a sample of tracks passing a set of \dEdx\ requirements
designed to enhance the fraction of charged kaons.
\end{itemize}

These samples were obtained and used as in Reference~\cite{muon} to obtain the 
fake rate uncertainty for identified muons from semileptonic decays 
in \Ztobb\ and \Ztocc\ events. 
This gave a multiplicative correction to the Monte Carlo simulation 
fake rate probability of $1.09\pm 0.10$.
The uncertainty on the correction factor strongly depends
on the particle composition of the hadron sample.

This procedure was repeated for the lower momentum spectrum 
and the particle composition applicable for this analysis 
to give a correction factor of $1.09\pm0.06$.
As the fraction of pions in this analysis is larger than in 
\cite{muon}, the contribution of the relatively large uncertainty
associated with the charged kaon sample becomes smaller and the overall
uncertainty on the fake
probability is reduced. Hence, the systematic uncertainty on the number
of hadrons that were mis-identified as muons is 154 events.\\

\noindent{\bf Electron identification efficiency:} 
The uncertainty on the simulated electron
identification efficiency was taken from Reference~\cite{Rbmain}. This
study is summarised below.

The most important variables that were
used in the electron identification neural network were 
the specific energy loss $\dEdx$, its error, 
and the ratio of the track's energy deposited in the calorimeter to
the track's momentum.
The Monte Carlo simulation and data distributions of these
variables were compared using samples 
of identified particles. 

The $\dEdx$ measurements were 
calibrated in data using samples of inclusive pions 
at low momenta and electrons at $45\thinspace\GeVc$ from Bhabha events. 
The quality of the calibration was checked with 
control samples, the most important of which were 
pions from $\PKs$ decays and electrons  from photons converting in the
detector. 
There was a smaller than $5\thinspace\%$ difference between the
mean $\dEdx$ measured in these samples in the data, and the
corresponding sample in the Monte Carlo simulation.
Similarly, the $\dEdx$ resolution in these 
samples have been studied and the data and Monte Carlo simulation 
were found to agree to within $8\thinspace\%$. 
The uncertainty on the electron identification efficiency from these two 
sources was found by varying both simultaneously, and was 
$\pm 2.5\thinspace\%$

A similar study has been performed for the
next most significant input variable $E/p$, 
which has a resolution in the Monte Carlo around
$10\thinspace\%$ worse than in the data. To correct for this 
the Monte Carlo has been reweighted to match the data, resulting 
in a variation of the efficiency of $\pm 2.7\thinspace\%$.

No significant contribution to the electron efficiency uncertainty
was found from the other input variables. 
The uncertainty arising from them was estimated from the 
statistical precision of these tests, which was less 
than $1\thinspace\%$ of the efficiency. 
In total, an uncertainty of $\pm 4.0\thinspace\%$ was assigned to the 
electron identification efficiency.\\

\noindent{\bf Electron mis-identification rate:} 
The uncertainty on the simulated electron fake rate
was evaluated as in \cite{Rbmain}. 
  The study used
  $\PKs$\to$\pi^+\pi^-$ decays and three-prong $\tau$ decays,
  and gave a fake rate uncertainty of $21\thinspace\%$ which
  corresponds to 17 events. \\

\noindent{\bf Photon conversion tagging efficiency:} 
  The uncertainty on the modelling 
   of the photon conversion tagging efficiency was estimated
  by comparing data and Monte Carlo samples of identified electrons
  with low momentum and low transverse momentum. 
  These samples had a very high electron purity and a photon
  conversion purity of 77\thinspace\%.
  The uncertainty on the photon conversion tagging efficiency was
  estimated to be $3.5\thinspace\%$ giving a systematic uncertainty of
   27 events.\\

\noindent{\bf Dalitz decays:}
 The uncertainty from this source arises from the modelling of the
 efficiency and from the limited knowledge of the $\pi^0$ and $\eta$ 
multiplicities.
A systematic
uncertainty was assigned for possible differences between the multiplicity of 
neutral pions and $\eta$ mesons in the Monte Carlo simulation and data. 
The production rates in the Monte Carlo were varied within the 
experimental uncertainties on the measured multiplicities~\cite{pdg},
giving a 1.2\thinspace\% uncertainty on the electron channel result.
Combining these sources of uncertainty, we estimated the uncertainty on
 the number of events from Dalitz decays to be 13 events.\\

\noindent{\bf Lepton transverse momentum:} To estimate the effect of
 the lepton transverse momentum criterion on the efficiency we
 compared the fraction of events that pass this criterion in data and
in the Monte Carlo simulation. We found this effect to be of the order
 of 1.3\%.

\subsection{{\boldmath \Dstar}\ Identification}
{\bf {\boldmath \Dstar}\ reconstruction efficiency:}
As the efficiency to
reconstruct a \Dstar\ was found to be smaller for low \Dstar\ momentum, the
compatibility of the data and the Monte Carlo \Dstar\ reconstruction 
efficiency  was compared by studying the ratio of the \Dstar\ yield in
Monte Carlo and data in different momentum regions. 
Since no significant
difference between low and high momentum regions was observed,
the statistical precision of this test was assigned as the systematic
uncertainty. 
The mis-modelling of $\Delta M $ resolution was also studied.
The observed difference in $\Delta M $ width between the data and
Monte Carlo was found to
be 0.15\thinspace\MeVcc\ resulting in a 
2.3\thinspace\% uncertainty on the \Dstar\
efficiency. 
The uncertainty arising from the modelling of $\dEdx$ was estimated by
comparing the selection to a modified version, where no $\dEdx$ requirements
where made. The change in the yield in the Monte Carlo and data 
was consistent. The 3.2\thinspace\% 
statistical precision of this test was assigned as
the uncertainty from this source.
Combining these effects, we obtained a 4.2\thinspace\% uncertainty on the \Dstar\
efficiency.\\

\noindent{ \bf {\boldmath \Dstar}\ background modelling:} The
background fitting
procedure was repeated changing the region of $\Delta M$ used for the
normalisation. In addition we also changed the \Dz\ mass sideband 
to include the lower sideband (1.26~\GeVcc $<~{\mathrm M_{K^-\pi^+}}~<$ 
1.56~\GeVcc).
Adding the differences due to the
above tests, and due to the uncertainties on the fitting parameters,
we obtained an uncertainty of 3.3\thinspace\% on the number of
\Dstar\ mesons.

\subsection{QCD and Fragmentation}

{\bf Secondary charm production modelling:} As there are no
measurements of the momentum spectrum of charmed hadrons from the
process of g\to\ccbar\ to test the modelling, the JETSET parameters 
with which the events were generated were varied according to Reference~\cite{Bill}.
The parton shower 
$\Lambda$ value was varied in the range 0.13 to 0.31\thinspace\GeV.  
The invariant mass cut-off of
the parton shower $\mathrm Q_0$ was varied between 1.4 and 2.5\thinspace\GeV.
The parameter $\sigma_q$, the width of the primary hadrons transverse
momentum distribution was varied in the range 
0.37 to 0.43\thinspace\GeV\
and $b$, the parameter of the Lund symmetric fragmentation function
was varied in the range 0.48 to 0.56\thinspace\GeV$^{-2}$. 
%
%
Full detector simulation was not available for all these variations of the 
model parameters, and therefore, estimates of their effect were made by applying 
appropriate cuts and smearing the event properties. 
In addition we have compared
HERWIG~\cite{HERWIG} and ARIADNE~\cite{ARIADNE} to JETSET.  
We have taken the largest
difference between the three predictions as an extra source of
uncertainty.\\
 

\noindent{\bf Fragmentation modelling:} 
The heavy-quark fragmentation was simulated using the
function of Peterson~\etal~\cite{peterson} and light-quark
fragmentation was simulated according to the Lund symmetric scheme.
The heavy-quark fragmentation model parameters
were varied to change the mean scaled
energy of weakly-decaying bottom and charm hadrons within their
experimental range: 
$\langle x_{E}\rangle_{\rm b}=0.702\pm 0.008$ and 
$\langle x_{E}\rangle_{\rm c}=0.484\pm 0.008$ respectively~\cite{LEPSLD}. 
In addition, the heavy-quark model was changed to the Lund symmetric model, 
to that suggested by Collins and
Spiller~\cite{CAS} and to that of Kartvelishvili~\etal~\cite{KAV} with parameters
tuned according to Reference~\cite{AJM}.
The largest difference between the Peterson fragmentation and the
other models was taken as a systematic uncertainty.

\noindent{\bf Jet scheme dependence:}
The stability of the results has been checked by changing the jet
finding algorithm to the Durham scheme, with $y_{\mathrm {cut}}$ set to 
$y_{\mathrm {cut}}= 0.015$; this choice of $y_{\mathrm {cut}}$ value was made 
as it optimised the significance of the observed signal. 
With this value, we obtained 
${\mathrm g^{\mathrm e}_{c\overline c}} = 0.0330\pm 0.0032$ and 
${\mathrm g^{\mu}_{c\overline c}} = 0.0401\pm 0.0043$ for the
electron and muon channels respectively
(here the uncertainties are statistical only). 
The slightly larger statistical uncertainties with respect to the Jade
jet finding results (equations 4 and 5) justify the choice of the Jade
algorithm for this analysis.
Considering the 57\thinspace\% overlap of the Jade and Durham samples,
the two results are consistent and no
additional systematic uncertainty was introduced.

\subsection{Heavy Quark Production and Decay}
{\bf Semileptonic decay modelling (lepton channels only):} 
Events with a prompt lepton were reweighted
as a function of the lepton momentum in the rest frame of
the decaying heavy hadron to simulate different models of semileptonic
decay as in \cite{LEPSLD}.
The semileptonic
decay model of Altarelli~\etal~\cite{ref:acm} (ACCMM), with
parameters tuned to CLEO data~\cite{ref:cleobl} for b decays and to
DELCO~\cite{ref:delco} and MARK~III~\cite{ref:markiii} data
for charm decays, was used for the central values, and was
combined with the \bD\ spectrum measured by CLEO~\cite{ref:cleobd} for
\bcl\ decays.
The model of Isgur~\etal~\cite{ref:isgw} (ISGW)
and their modified model (\ISGWSS) with
the fraction of \Dss\ decays determined from CLEO data~\cite{ref:cleobl}
were used to determine the systematic
uncertainty due to the ${\mathrm{b}}\rightarrow\ell$ spectrum. \\

\noindent{\bf Charm and bottom branching ratios:}  The dependence on the 
semileptonic branching ratios, ${\mathrm{b}}\rightarrow\ell$, 
${\mathrm{c}}\rightarrow\ell$, ${\mathrm{b}}\rightarrow{\mathrm
c}\rightarrow\ell$, 
 as well as
the hadronic branching ratios of $\mathrm b$\to\Dstar, $\mathrm c$\to
\Dstar, ${\mathrm D^{\star +}}$\to ${\mathrm D^0} \Pip$ 
and ${\mathrm D^0}$\to \Km\Pip, has 
been investigated by varying them within their experimental uncertainties
\cite{pdg,LEPSLD,ARGUS,OPALcsl}. 

A possible energy dependence of the B$(\mathrm c$\to\Dstar)$\times$ 
B(\Dstar\to ${\mathrm D^0}\pi)$ rate has been taken into
account by comparing the average of LEP measurements,
$0.1527\pm0.0092$
\thinspace\cite{LEPSLD}, with
an average of lower energy measurements,
$0.184\pm0.014$\thinspace\cite{pdg}. 
Typical \Dstar\ energies in this analysis were between these extremes, and the 
uncertainty on this product branching ratio has been inflated  
to 0.015 to account for these possible effects.
In contrast, the OPAL and ARGUS \cite{OPALcsl,ARGUS} 
values for the semileptonic
branching ratio $\mathrm c\rightarrow \ell$ are consistent, and no additional 
uncertainty has been assigned.\\


\noindent{\bf Partial hadronic widths:} The partial hadronic widths of
bottom and charm with which
both the efficiency
and the jet mis-assignment background were estimated, were taken
from a combination of LEP and SLD results~\cite{LEPSLD}. The error on
the partial widths is a source of systematic uncertainty on
\gcc\ \\

\noindent{\bf g{\boldmath\to\bbbar}:} The uncertainty on the average 
measured value of \gbb\
gave an uncertainty of 14, 15, and 1.1 events  
on the number of background events for the electron, muon, 
and \Dstar\ channels respectively.\\

\noindent{\bf Jet mis-assignment background modelling:}
This background could be mis-modelled if the efficiency for tagging a
heavy quark jet was incorrect, 
or if the fraction of primary heavy quark jets which
were identified as the gluon jet candidates was mis-modelled in the 
Monte Carlo.
We tested these cases, using enriched \Ztobb\ samples 
(as described in section~\ref{sec:LeptonAnalysis:datamc}), 
As data and Monte Carlo were consistent in all cases
we assigned the statistical precision of these tests as a systematic
uncertainty. We assigned the difference solely to the signal, using the
test purity, and took the larger difference as a systematic
uncertainty. 
 \\

\noindent{\bf Jet mass criterion:} The systematic uncertainty
associated with the jet mass criterion was found by comparing the
fraction of events passing this selection in data and Monte Carlo.
The systematic uncertainty assigned to this source is 2.3\%.\\

\noindent{\bf B hadron decay product multiplicity:} The artificial neural
  network used to reject \Ztobb\ events was sensitive to the B hadron
  charged track multiplicity. Consequently \Ztobb\ events were
  reweighted to reproduce
  the experimental uncertainty on the measured 
  multiplicity~\cite{LEPSLD}. The uncertainty on the number
  of background events rejected associated with this procedure was 
  estimated to be 5\thinspace\%.

\subsection{Other sources}
\noindent{\bf Detector modelling:}
The resolution of the central tracking in the Monte Carlo had an
effect on the predicted efficiencies and background.
The simulated resolutions were varied by $\pm 10\thinspace\%$
relative to the values that optimally describe the
data following the studies in~\cite{Rbmain}.
The analysis was repeated and the efficiencies and background
estimation were recalculated.\\

\noindent{\bf Monte Carlo Statistics:}
This was the uncertainty due to
the finite size of the Monte Carlo samples used to determine the
efficiencies and background.
 
\begin{table}
\begin{center}
\begin{tabular}{|l|c|c|c|} \hline
 & &  & \\*[-11pt]
{\bf Source of uncertainty} & 
$\delta(\gccd^{\mathrm e})/\gccd^{\mathrm e}$ (\%)& 
$\delta(\gccd^\mu)/\gccd^\mu$ (\%)& 
$\delta(\gccd^{\mathrm D^*})/\gccd^{\mathrm D^*}$ (\%)\\
\hline
Lepton efficiency & 4.9 & 5.8 &-\\ 
Lepton mis-identification & 2.5 & 18.4 & - \\ 
Efficiency of photon conversion tagger & 2.7 & - & -\\
Dalitz decay multiplicities and efficiencies & 1.4 &- &- \\
Lepton transverse momentum & 1.4 & 1.2 & - \\
\hline
\Dstar\ selection & - & - & 7.1 \\ 
\Dstar\ background modelling & - & - & 5.7 \\ 
\hline
$\Lambda_{QCD}$  & 2.6 &2.6 &2.6\\ 
Parton shower mass cut-off & 0.2 & 0.2 & 0.2 \\ 
Primary hadron trans. mom. width ($\sigma_{\mathrm q}$)  & 1.5 & 1.5 & 1.5 \\
Lund symmetric fragmentation parameter $b$ & 1.8 & 1.8 & 1.8  \\
JETSET - HERWIG - ARIADNE & 4.1 & 4.1 & 4.5\\ \hline
$\langle x_{E}\rangle_{\rm b}=0.702\pm 0.008$ & 1.1 & 1.0 & 2.7 \\ 
$\langle x_{E}\rangle_{\rm c}=0.484\pm 0.008$ & 1.3 & 1.3 & 3.5 \\ 
Heavy quark fragmentation model & 1.3 & 1.1 & 4.3 \\
\hline
$\mathrm b\rightarrow \ell$ model  & 0.1 & 0.1 &-\\ 
$\mathrm c\rightarrow \ell$ model & 0.4 & 0.4 &- \\ 
${\mathrm b}\rightarrow {\mathrm c}\rightarrow\ell$ model &0.2 &0.2 &-\\
B( $\mathrm b\rightarrow \ell$)=(10.99\plm0.23)\thinspace\%& 0.4 & 0.4 & - \\
B($\mathrm c\rightarrow \ell$)=(9.5\plm0.7)\thinspace\%& 7.8 & 7.6 & -\\
B( ${\mathrm b}\rightarrow {\mathrm c}\rightarrow\ell$)=(7.8\plm0.6)\thinspace\%
& 1.1 & 1.0 &- \\
B($\mathrm b$\to\Dstar)=(22.7\plm1.6)\thinspace\% & - & - & 1.5 \\
B( $\mathrm c$\to\Dstar)$\times$ B(\Dstar\to ${\mathrm D^0}
\pi$)=(15.3\plm1.5)\thinspace\% &- &- &7.0 \\
B(${\mathrm D^0}$\to K$\pi$)=(3.85\plm0.09)\thinspace\%& - & - & 2.1 \\
$\GbbGhad$=0.2170\plm0.0009 & 0.3 & 0.3 & 0.1 \\ 
$\GccGhad$=0.1734\plm0.0048 & 1.4 & 1.2 & 0.6 \\ 
g\to\bbbar=(2.69\plm0.67)$\times10^{-3}$ & 1.8 & 1.5 & 1.5 \\ 
Jet mis-assignment & 1.7 & 1.5 & 0.6 \\
Jet mass cut & 2.3 & 2.3 & - \\
B hadron decay multiplicity &3.6  & 3.2 & 1.3 \\ 
\hline
Detector resolution & 2.5 & 1.9 & 6.1\\ 
Monte Carlo statistics  & 2.9 & 2.7 & 5.7 \\
\hline \hline
{\bf Total systematic uncertainty}     &{\bf 13.3} &{\bf 22.2} &{\bf 16.8}\\
\hline
\end{tabular}
\caption{Summary of the systematic uncertainties on the measured \gcc\ values.}
\label{tab:sys}
\end{center}
\end{table}

\section{Results and Conclusions}
\label{sec:summary}
The three measurements of the rate of secondary charm quark
production are
\begin{eqnarray}
\gccd^{\mathrm e}& =   0.0303\pm 0.0028\pm 0.0040,\\
\gccd^\mu& =           0.0353\pm 0.0037\pm 0.0078,\\
\gccd^{\mathrm D^*}& = 0.0408\pm 0.0122\pm 0.0069,
\end{eqnarray}
where the first uncertainty is statistical and the second systematic.
By averaging the leptonic channels and taking into account correlations in
the systematic uncertainties we obtain
\begin{equation}
\gccd^\ell = 0.0311\pm 0.0022\pm 0.0041,
\end{equation}
and combining this with the hadronic channel gives
\begin{equation}
\gccd = 0.0320\pm 0.0021\pm 0.0038.
\end{equation}
These results can be compared with the previous OPAL 
measurements\thinspace\cite{Dspm,gcc}. As 
the values used for the rate of the process g\to\bbbar\ and 
the branching ratio B($\mathrm c\rightarrow \ell$) have changed since 
the previous publication, a comparison should be made after
correcting for these effects. The old lepton results \cite{gcc} should be
scaled by a factor of 1.09. 
Thus the scaled results are
$\gccd^{\mathrm e} =   0.0238\pm 0.0033\pm 0.0045$,
$\gccd^\mu =  0.0309\pm 0.0063\pm 0.0111$ and 
$\gccd^{\mathrm D^*} = 0.044\pm 0.014\pm 0.015$.
The analyses described in this document
use a larger data sample and 
different selection criteria than before. In particular, the different
$y_{\mathrm cut}$ value and lepton momentum range resulted in a small
overlap of the present data sample and the previous analysis sample.
With these changes, less
than one third of the current electron analysis data sample
overlaps the previous electron analysis. The increased momentum range
of muons in this analysis resulted in having only 15\thinspace\% of this
muon sample common to the previous analysis. The previous \Dstar\ analysis
\cite{Dspm} used only 1.25 million hadronic \Zz\ decays. Therefore, 
the number of common event to the present \Dstar\ analysis is lower
than 30\thinspace\%. Taking into account these limited overlaps, the present
and previous results are statistically consistent.

In recent years, the OPAL data calibration has also been refined and the
Monte Carlo simulation of the detector response had improved
significantly. 
In view of these changes, the 
analysis presented in this paper is more accurate than the previous 
analysis in both the statistical significance and the proper description 
of the experimental environment. In conclusion, although the two
results are statistically largely
independent, they are dominated by systematic uncertainties, which are
better understood for the new measurement. This result therefore
supersedes the previous OPAL measurement.

The result presented in this paper, \gcc=0.0320\plm 0.0021\plm 0.0038,
is higher than all the theoretical predictions. In particular, the most recent
prediction (\gcc=2.007\thinspace\%~\cite{miller}), which is higher
than most other predictions, is 2.7 standard
deviations below the result presented in this paper.

\section*{Acknowledgements:}

We particularly wish to thank the SL Division for the efficient operation
of the LEP accelerator at all energies
 and for their continuing close cooperation with
our experimental group.  We thank our colleagues from CEA, DAPNIA/SPP,
CE-Saclay for their efforts over the years on the time-of-flight and trigger
systems which we continue to use.  In addition to the support staff at our own
institutions we are pleased to acknowledge the  \\
Department of Energy, USA, \\
National Science Foundation, USA, \\
Particle Physics and Astronomy Research Council, UK, \\
Natural Sciences and Engineering Research Council, Canada, \\
Israel Science Foundation, administered by the Israel
Academy of Science and Humanities, \\
Minerva Gesellschaft, \\
Benoziyo Center for High Energy Physics,\\
Japanese Ministry of Education, Science and Culture (the
Monbusho) and a grant under the Monbusho International
Science Research Program,\\
Japanese Society for the Promotion of Science (JSPS),\\
German Israeli Bi-national Science Foundation (GIF), \\
Bundesministerium f\"ur Bildung, Wissenschaft,
Forschung und Technologie, Germany, \\
National Research Council of Canada, \\
Research Corporation, USA,\\
Hungarian Foundation for Scientific Research, OTKA T-029328, 
T023793 and OTKA F-023259.\\



\begin{thebibliography}{99}
\bibitem{seymour}
     M.H.~Seymour,  Nucl. Phys. {\bf B 436} (1995) 163.
\bibitem{nason}
        M.L.~Mangano, P. Nason, Phys. Lett. {\bf B 285} (1992) 160.
\bibitem{seyi}
        M.H. Seymour, \ZPhys\  {\bf C 63} (1994) 99.
\bibitem{miller} D.J. Miller, M.H. Seymour, Phys. Lett. {\bf B 435} (1998) 213.
\bibitem {Dspm}
     \OPALColl, R. Akers \etal, \ZPhys\ {\bf C 67} (1995) 27.
\bibitem {gcc}
     \OPALColl, R. Akers \etal, Phys. Lett. {\bf B 353} (1995) 595.
\bibitem {gbb1}ALEPH Collab., R. Barate \etal, Phys. Lett. {\bf B
     434} (1998) 437.
\bibitem {gbb2} DELPHI Collab., P. Abreu \etal,
                                    Phys. Lett. {\bf B 405} (1997) 202.
\bibitem{OPAL}
    \OPALColl, K. Ahmet \etal, \NIM\ {\bf A 305} (1991) 275;\\
    \OPALColl, P.P. Allport \etal, \NIM\ {\bf A 324} (1993) 34;\\
    \OPALColl, P.P. Allport \etal, \NIM\ {\bf A 346} (1994) 476.
\bibitem{evsel}
    \OPALColl, G. Alexander \etal, \ZPhys\ {\bf C 52} (1991) 175.
\bibitem{seventhree}
T. Sj\"ostrand, Comp. Phys. Comm. {\bf 82} (1994) 74;\\
T. Sj\"ostrand, Comp. Phys. Comm. {\bf 39} (1986) 347;\\
M. Bengtsson and T. Sj\"ostrand, Comp. Phys. Comm. {\bf 43} (1987) 367;\\
T. Sj\"ostrand, Int. J. of Mod. Phys. {\bf A 3} (1988) 751.
\bibitem{peterson} C. Peterson \etal,  Phys. Rev. {\bf D 27} (1983) 105.
\bibitem{LEPSLD} The LEP Collaborations, ALEPH, DELPHI, L3 and
  OPAL, Nucl.~Instr. and~Meth. {\bf A 378} (1996) 101;
Updated averages are given in `{\it Input Parameters for the LEP 
Electroweak Heavy Flavour Results for Summer 1998 Conferences'}, LEPHF
 98-01 (see http://www.cern.ch/LEPEWWG/heavy/); {\it A Combination of 
Preliminary Electroweak Measurements and Constraints on the Standard Model,} 
 ALEPH, DELPHI, L3 and OPAL collaborations, the LEP Electroweak
     Working Group and the SLD Heavy Flavour and
    Electroweak Groups, 
    CERN-EP/99-015.
\bibitem{pdg} The Particle Data Group, C. Caso \etal, Eur. Phys. J. {\bf C 3} (1998) 1. 
\bibitem{gopal} J. Allison \etal, \NIM\ {\bf A 317} (1992) 47.
\bibitem{otherjets}
     \JADEColl, W. Bartel \etal, \ZPhys\ {\bf C 33} (1986) 23;\\
      \JADEColl, S. Bethke \etal, \PhysLett\ {\bf B 213} (1988) 235;\\
    S. Bethke \etal, Nucl. Phys. {\bf B 370} (1992) 310;\\ 
        \OPALColl, R. Akers \etal, Z. Phys. {\bf C 63} (1994) 197.
\bibitem{Rbmain} \OPALColl, G. Abbiendi \etal, 
Eur. Phys. J. {\bf C 8} (1999) 217.
\bibitem{muon}
        \OPALColl, R. Akers \etal, Z. Phys. {\bf C 60} (1993) 199.
\bibitem{VNN} \OPALColl, R. Akers \etal, Z. Phys. {\bf C 66} (1995) 19.
\bibitem{OPALcsl} \OPALColl, G. Abbiendi \etal, Eur. Phys. J. {\bf C
        8} (1999) 573.
\bibitem{ARGUS} ARGUS Collab., H. Albrecht \etal, Phys Lett {\bf B 374} (1996) 249.
\bibitem{Bill}\OPALColl, G. Alexander \etal, Z. Phys {\bf C 69} (1996) 543.
\bibitem{HERWIG} G. Marchesini, B.R. Webber \etal, Comp. Phys. Comm. 
{\bf 67} (1992) 465.
\bibitem{ARIADNE} L. Lonnblad, Comp. Phys. Comm. {\bf 71} (1992) 15.
\bibitem{CAS} P. Collins and T. Spiller, J. Phys. {\bf G 11} (1985) 1289.
\bibitem{KAV} V.G. Kartvelishvili, A.K. Likehoded and V.A. Petrov,
     Phys. Lett. {\bf B 78} (1978) 615.
\bibitem{AJM} \OPALColl, G. Alexander \etal, Phys. Lett. {\bf B 364} (1995) 93;\\
       \OPALColl, G. Alexander \etal, Z. Phys. {\bf C 72} (1996) 1.
\bibitem{ref:acm}
    G.Altarelli \etal, Nucl. Phys. {\bf B 208} (1982) 365.
\bibitem{ref:cleobl}
     CLEO Collab., S. Henderson \etal,
     Phys. Rev. {\bf D 45} (1992) 2212. 
\bibitem{ref:delco} DELCO Collab., W. Bacino \etal,
    Phys. Rev. Lett. {\bf 43} (1979) 1073.
\bibitem{ref:markiii} MARK III Collab.,
R.M. Baltrusaitis \etal, Phys. Rev. Lett. {\bf 54} (1985) 1976.
\bibitem{ref:cleobd}
   CLEO Collab., D. Bortoletto \etal, Phys. Rev. {\bf D 45} (1992) 21.
\bibitem{ref:isgw}
    N. Isgur, D. Scora, B. Grinstein and M. Wise,
    Phys. Rev. {\bf D39} (1989) 799.
\end{thebibliography}
\end{document}